\documentclass{article}
\usepackage{dcase2022,amsmath,graphicx,url,times,booktabs, tabularx}
\usepackage{float}

\title{An investigation on selecting audio pre-trained models for audio captioning}

 \name{Peiran Yan$^{1}$,
       Shengchen Li$^{1}$
       }

 \address{$^1$ Xi’an Jiaotong-Liverpool University, Suzhou, China
  }

\begin{document}

\ninept
\maketitle

\begin{sloppy}

\begin{abstract}
Audio captioning is a task that generates description of audio based on content. Pre-trained models are widely used in audio captioning due to high complexity. Unless a comprehensive system is re-trained, it is hard to determine how well pre-trained models contribute to audio captioning system. To prevent the time consuming and energy consuming process of retraining, it is necessary to propose a preditor of performance for the pre-trained model in audio captioning. In this paper, a series of pre-trained models are investigated for the correlation between extracted audio features and the performance of audio captioning. A couple of predictor is proposed based on the experiment results.The result demonstrates that the kurtosis and skewness of audio features extracted may act as an indicator of the performance of audio captioning systems for pre-trained audio due to the high correlation between kurtosis and skewness of audio features and the performance of audio captioning systems.

\end{abstract}

\begin{keywords}
Audio caption, SPIDER, Kurtosis, encoder, skewness
\end{keywords}

\section{Introduction}
\label{sec:intro}
By applying natural language description, the automated audio captioning (AAC) system, as a cross-modal translation task containing the techniques of natural language processing and audio processing, describes an audio file \cite{drossos2017automated}. The visual perception’s objects are defined by its color, size, shape and its spatial position to other objects. In contrast, auditory perception focus on the audio events and its corresponding physical properties, its relationship with other events, and interim information of those audio events \cite{xu2021investigating}. Audio captioning could be applied in several areas, which includes helping the hearing-impaired to realize environmental sounds, investigating sounds in smart cities for safety surveillance, and subtitling for sound in a television show \cite{mei2021encoder}. 

Kurtosis is the scale-free and location-free movement of the probability mass from the shoulders of the distribution to their tails and center \cite{balanda1988kurtosis}. With respect to different forms of kurtosis, it has the dissimilar positionings and scaling techniques of the shoulders result \cite{booker1998brief}. Kurtosis often used to describe the shape feature of a distribution. For the audio feature in vector representation, kurtosis could verify whether the data sets are light-tailed or heavy-tailed compared with normal distribution. Specifically, data sets have heavy tails or outliers with high kurtosis. Data sets have light tails or lack of outliers with low kurtosis. The theory of kurtosis is also applied in tests of normality and research of robustness to normal theory procedures, such as in Wilcox \cite{wilcox1990comparing}. 

In a set of data, skewness is quantified as the degree of an asymmetry varies from a normal distribution. The curve is said to be skewed if it is moved to the right or to the left.Skewness is also significant to the deliberation on the characteristics of large disasters in clarify the equity risk payment \cite{mehra1988equity}. For symmetric distribution (normal distribution), it has zero skewness. The asymmetric distribution with a longer left tail holds negative skewness, and the distribution with the tail to right holds positive skewness \cite{brys2003comparison}. 

In \cite{xu2021investigating}, it states that the current mainstream of training architecture AAC is the end-to-end encoder-decoder architecture, which the captions are offered as the supervision label to the audio files. In the encoder-decoder audio captioning system, the audio encoder extracts the abstract audio feature from the dataset. Then the text decoder forecast the words based on the extracted audio feature. Because of encoding all the above-mentioned multidimensional information from audio files without the explicit supervision, the level of complexity for AAC encoder training is improved \cite{xu2021investigating}. Thereby, it is significant to describe the hierarchical structure of the audio topis. The “CNN-Transformer” framework of encoder-decoder architecture was shown to good performance in DCASE 2021 Task 6, which ranked 3rd over all submissions \cite{mei2021encoder}. Thereby, this proposed system is selected to run experiments of relationship between kurtosis, skewness and performance (the architecture of selected system is shown in figure 1).

		\begin{figure}
        \centering
         \includegraphics[height=2.5in ,width=2.3in]{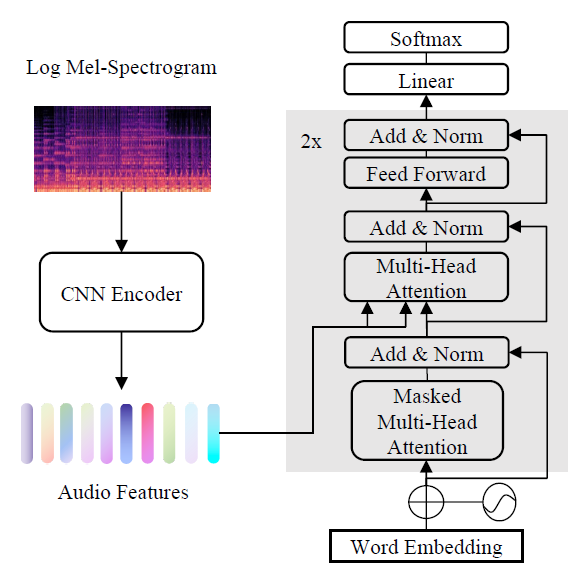}
        \caption{Architecture of the selected system \cite{mei2021encoder}}
       \label{fig:1}
       \end{figure}

Training an end-to-end cross-modal audio captioning system without pre-trained model is complicated and random because of the amount of data in audio captioning tasks is restricted. To solve this issue, transfer learning is applied in this proposed system to improve the amount of parameters. Specifically, the pre-trained audio models from related audio tasks such as sound event detection and audio tagging are applied in audio encoders to improve the ability of extracting audio feature \cite{chen2020audio}. To make the results of experiments more comprehensive and accurate, this experiment uses three types of neuron network for audio encoders, which are CNN10, CNN6, ResNet22. All of these neuron networks are published in PANNs \cite{kong2020panns}. During the training procedure, other settings are the same and pretrained-models are applied for all audio encoders. The system will record the value of kurtosis and skewness of extracted audio feature from audio encoder and the scores of SPIDER to check the relationship between them for each epoch. 

Another problem is how to set the termination of training for the proposed system. The previous work usually set the number of epoch by testing it randomly and provide a specific number. In this experiment, the score of SPIDEr is improved with the improvement of kurtosis and skewness of extracted audio feature from encoder. At the back stage of training, the value of kurtosis and skewness become stable which not vary drastically. But the score of SPIDEr become stable of lower than before. Therefore, the system could terminate the training procedure after the value of kurtosis and skewness not vary for serval epoch.  

The residual sections of this paper are arranged as below. In Section 2, it will describe the details of selected systems and different tested audio encoders. Section 3 will demonstrate the experimental setup of calculating kurtosis and skewness of extracted audio feature from encoders. Experiments results and analysis are shown in Section 4. Eventually, the conclusion of this work will be in Section 5.

\section{PROPOSED METHOD}
\label{sec:format}
The selected system contains of a audio encoder (CNN10, or CNN6, or ResNet22) and a Transformer decoder. The encoder receives the log mel-spectrogram $X$ $\in$ $R^{T x D}$ of the audio files as input feature and makes feature vector $v$ $\in$ $R^{T' x D'}$. According to the feature vector $v$ that extracted from encoder and prior words $w_0$ to $w_{n-1}$, the Transformer decoder predicts the posterior probability of the n-th word  $w_n$. 

\subsection{Encoders}
Convolutional neural networks (CNNs) have demonstrated influential ability in extracting audio features and have been broadly applied in audio processing related tasks \cite{cruciani2020feature}. Residual neural networks (RANNs) have shown more accurate and efficient in some audio tasks and the system has less computational complexity \cite{verbitskiy2021eranns}. Thereby, this experiment selects CNN10, CNN6, ResNet22 proposed in the pre-trained audio neuron networks (PANNs) \cite{kong2020panns} is applied as the encoder to investigate the relationship between kurtosis and skewness of extracted audio feature and the performance of system. 

\subsubsection{CNN10}
\begin{center}
 \begin{tabular}{||c||} 
 \hline
 CNN10  \\ [0.5ex] 
 \hline\hline
 (3 $\times$ 3 \@ 64,\\BN, ReLU) $\times$ 2 \\ 
 \hline
 Pooling 2 $\times$ 2  \\
 \hline
 (3 $\times$ 3 \@ 128,\\BN, ReLU) $\times$ 2\\
 \hline
 Pooling 2 $\times$ 2 \\
 \hline
 (3 $\times$ 3 \@ 256,\\BN, ReLU) $\times$ 2\\ 
 \hline
 Pooling 2 $\times$ 2\\
 \hline
 (3 $\times$ 3 \@ 512,\\BN, ReLU) $\times$ 2\\ 
 \hline
 Global Pooling\\
 \hline
 FC 512, ReLU\\
 \hline
\end{tabular}
\end{center}

A simple 10-layer CNN is proposed in the pre-trained audio neuron networks (PANNs) \cite{kong2020panns}. The above is the architecture of CNN10. 

\subsubsection{CNN6}
\begin{center}
 \begin{tabular}{||c||} 
 \hline
 CNN6  \\ [0.5ex] 
 \hline\hline
 5 $\times$ 5 \@ 64\\
  BN, ReLU \\ 
 \hline
 Pooling 2 $\times$ 2  \\
 \hline
 5 $\times$ 5 \@ 128\\
  BN, ReLU \\
 \hline
 Pooling 2 $\times$ 2 \\
 \hline
  5 $\times$ 5 \@ 256\\
  BN, ReLU \\
 \hline
 Pooling 2 $\times$ 2\\
 \hline
  5 $\times$ 5 \@ 512\\
  BN, ReLU \\ 
 \hline
 Global Pooling\\
 \hline
 FC 512, ReLU\\
 \hline
\end{tabular}
\end{center}

A simple 6-layer CNN is proposed in the pre-trained audio neuron networks (PANNs) \cite{kong2020panns}. The above is the architecture of CNN6. 

\subsubsection{ResNet22}
\begin{center}
 \begin{tabular}{||c||} 
 \hline
 ResNet22  \\ [0.5ex] 
 \hline\hline
 (3 $\times$ 3 \@ 512, BN, ReLU) $\times$ 2\\ 
 \hline
 Pooling 2 $\times$ 2  \\
 \hline
 (BasicB \@ 64) $\times$ 2 \\
 \hline
 Pooling 2 $\times$ 2 \\
 \hline
  (BasicB \@ 128) $\times$ 2\\
 \hline
 Pooling 2 $\times$ 2\\
 \hline
  (BasicB \@ 256) $\times$ 2\\ 
 \hline
  Pooling 2 $\times$ 2\\
 \hline
 (BasicB \@ 512) $\times$ 2\\
 \hline
  Pooling 2 $\times$ 2\\
 \hline
 (3 $\times$ 3 @ 2048, BN, ReLU) $\times$ 2\\
 \hline
 Global Pooling\\
 \hline
 FC 2048, ReLU\\
 \hline
\end{tabular}
\end{center}

A simple 22-layer RANN is proposed in the pre-trained audio neuron networks (PANNs) \cite{kong2020panns}. The above is the architecture of ResNet22. 

\subsection{Decoder}
The decoder of selected system contains three elements, a word embedding layer, a standard Transformer decoder and a linear layer. Firstly, through the word embedding layer, every input word is firstly encoded to a vector of constant dimension. Showing state-of-the-art performance and manipulating sequential data are the functionalities of Transformer in the area of natural language processing \cite{vaswani2017attention}. The audio features extracted from encoder and the word embeddings from word embedding layer are transformed to the Transformer decoder. The Transformer decoder applied in selected system contains two transformer decoder blocks with four heads. The dimension of the hidden layer is 128. The linear layer aims to output the probability distribution along the vocabulary.

\subsection{Transfer Learning}
The pre-trained audio models could be applied in encoder when the encoder extracts audio features from audio files. To extract audio features, some pre-trained audio models had been proposed. PANNs \cite{kong2020panns}, as one of the models of audio tagging task pre-trained on the AudioSet \cite{gemmeke2017audio} dataset, shows state-of-the-art performance in many related audio tasks. All encoders (CNN10, CNN6, ResNet22) are used pre-trained models in PANNs to initialize the parameters in the encoder.

\section{EXPERIMENTS}
\label{sec:pagelimit}

\subsection{Dataset}
Clotho \cite{drossos2020clotho}, as one of audio captioning datasets, which the audio files are all gathered from online platform of Freesound archive. The length of audio files ranges randomly from 15 to 30 seconds. Every audio file has five captions which are the descriptions of audio files. All the captions range randomly from 8 to 20 words. In this experiment, the development set and validation set are merged together to give a new training set. For training procedure, the five captions and its audio clip are merged together as a training sample. For evaluation procedure, all 5 ground-truth captions of an audio file are used to compare with the forecasted caption for metric computation. This experiment will use Clotho to investigate the relationship between kurtosis and skewness of extracted audio feature and the performance of system. 

\subsection{Data pre-processing}
64-dimensional log mel-spectrograms are applied as the input features which hold a 1024-point Hanning window with a hop size of 512-points. The data augment method in this experiment is SpecAugment \cite{park2019specaugment}, which manipulates on input features of the audio clip by applying time masking and frequency masking. For captions, all captions in Clotho and AudioCaps are transferred to lower case with punctuation eliminated. For each caption, the beginning and end of each caption pad two special tokens “<sos>” and “<eos>”. The vocabularies of Clotho and AudioCaps are combined together that hold a vocabulary including 6636 words. 

\subsection{Experimental setups}
All parameters of training and evaluation are same as the selected system's original setting, which is shown in \cite{mei2021encoder}. The batch size is 12. The encoders of CNN10, CNN6, ResNet22 are the variables of this experiment. All experiments are set the investigate the relationship between kurtosis and skewness of extracted feature and score of evaluation metrics of system by using different encoders. After the feature extracted from the encoder, the system will calculate the kurtosis of the feature and record it. The feature from the encoder are 3-dimension, which are $time x batch x channel$. Firstly, the system calculate the kurtosis of $channel$, and append it into corresponding batch list. Finally, calculate the mean of each batch, which the value is the final value. Mathematically, the formulas to calculate the kurtosis: \begin{equation}K=\frac{\frac{1}{n} \sum\limits_{i=1}^{n}\left(x_{i}-\bar{x}\right)^{4}}{\left(\frac{1}{n} \sum\limits_{i=1}^{n}\left(x_{i}-\bar{x}\right)^{2}\right)^{2}}\end{equation}.

For the quation 1, K is the kurtosis, n is the number of items. This experiments use the scipy.stats python library to calculate the kurtosis. 

calculate the mean of each batch, which the value is the final value. Mathematically, the formulas to calculate the skewness:
\begin{equation}\tilde{\mu}_{3}=\frac{\sum_{i}^{N}\left(X_{i}-\bar{X}\right)^{3}}{(N-1) * \sigma^{3}}\end{equation}

For the equation 2, it is the skewness, N is the number of items. This experiments use the scipy.stats python library to calculate the skewness.

\subsection{Evaluation metrics}
The audio captioning systems are evaluated by captioning metrics (CIDEr, SPIDEr and SPICE) and machine translation metrics ($BLEU_{n}$, METEOR, and $ROUGE_{l}$). CIDEr \cite{vedantam2015cider} uses term frequency inverse document frequency (TF-IDF) weights to n-grams and estimates the cosine similarity between it. SPICE \cite{anderson2016spice} converts captions to scene graphs and calculates F-score according to tuples in it. SPIDEr \cite{liu2017improved} is a linear grouping of CIDEr and SPICE, the CIDEr score confirms captions are syntactically fluent, while the SPICE score confirms captions are semantically dependable to the audio file. $BLEU_{n}$ \cite{papineni2002bleu} is estimated as a weighted geometric average of altered accuracy of n-grams. METEOR \cite{agarwal2007meteor} calculates a harmonic average of accuracy and recall according to word level matches between references and the candidate sentence.  $ROUGE_{l}$ \cite{lin2004rouge} estimates F-measures according to the longest common subsequence. In this experiment, the system will mainly focus the SPIDEr, which ranks the submission based on SPIDEr in DCASE 2021 Task6.

\section{Results}
\label{sec:pagestyle}
The score of SPIDEr is evaluated on evaluation set of Clotho. The figures below are the relationship between kurtosis and skewness of extracted feature the score of SPIDEr during training. 
		\begin{figure}[H]
        \centering
         \includegraphics[height=1.5in ,width=2.3in]{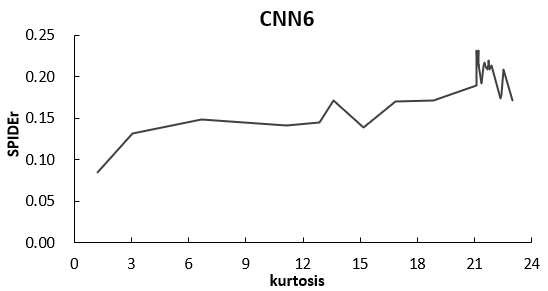}
        \caption{Kurtosis of Encoder: CNN6}
       \label{fig:1}
       \end{figure}
      
      		\begin{figure}[H]
        \centering
         \includegraphics[height=1.5in ,width=2.3in]{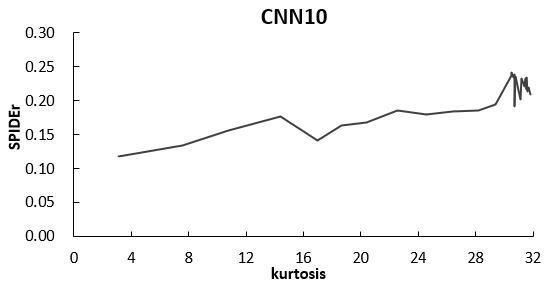}
        \caption{Kurtosis of Encoder: CNN10}
       \label{fig:1}
       \end{figure}

            \begin{figure}[H]
        \centering
         \includegraphics[height=1.5in ,width=2.3in]{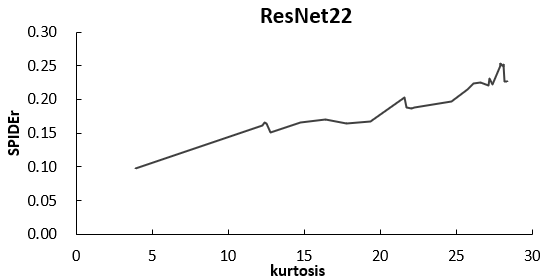}
        \caption{Kurtosis of Encoder: ResNet22}
       \label{fig:1}
       \end{figure}
       
		\begin{figure}[H]
        \centering
         \includegraphics[height=1.5in ,width=2.3in]{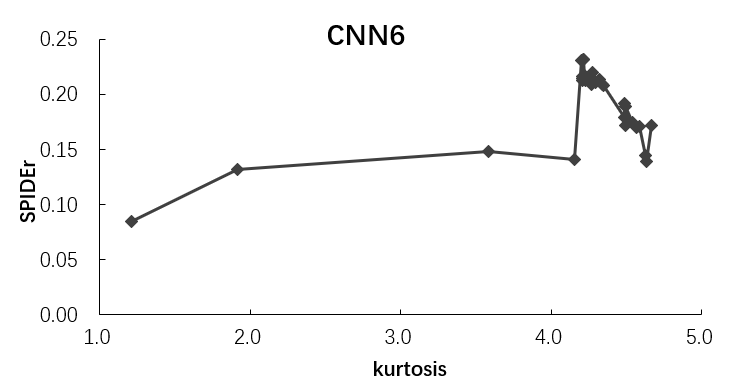}
        \caption{Skewness of Encoder: CNN6}
       \label{fig:1}
       \end{figure}
       
       	\begin{figure}[H]
        \centering
         \includegraphics[height=1.5in ,width=2.3in]{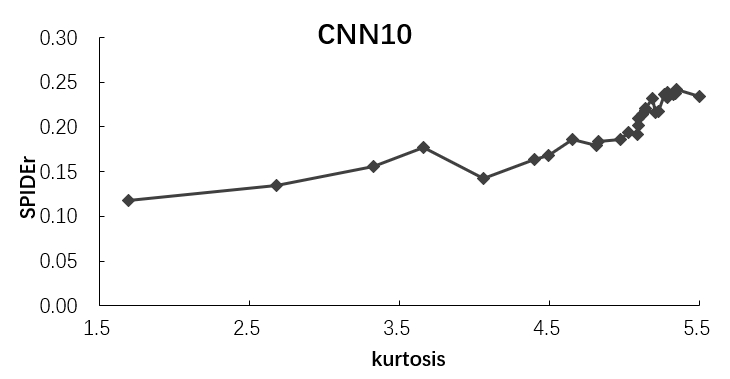}
        \caption{Skewness of Encoder: CNN10}
       \label{fig:1}
       \end{figure}
       
        \begin{figure}[H]
        \centering
         \includegraphics[height=1.5in ,width=2.3in]{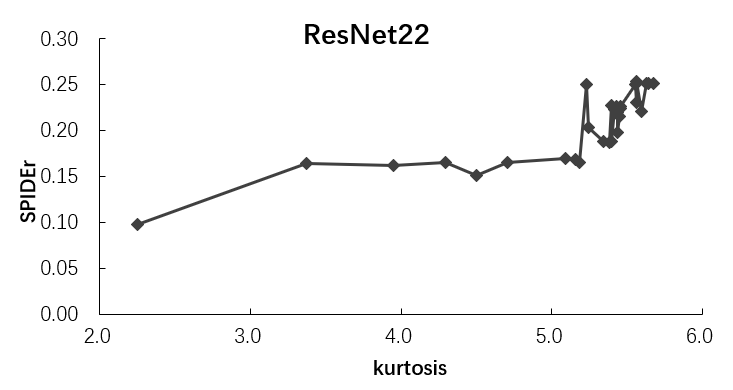}
        \caption{Skewness of Encoder: ResNet22}
       \label{fig:1}
       \end{figure}

The figures 2, 3, 4 shown that the relationship between kurtosis of extracted audio feature from encoders (CNN6, CNN10, ResNet22) and the score of SPIDEr in evaluation set of Clotho. The figures 5, 6, 7 shown that skewness. It is obvious that with the improvement of kurtosis and skewness of extracted feature, the score of SPIDEr is also improved generally. In the future, we could select the pre-trained model with the higher kurtosis and skewness.

\section{Conclusion}
This paper had presented how the kurtosis and skewness of extracted feature from encoder matches score of SPIDEr of “Encoder+Transformer” audio captioning system. The experiments are made in CNN6, CNN10, ResNet22 encoders to reduce the contingencies. All the results shows that the score of SPIDEr is improved with the improvement of kurtosis and skewness of extracted audio feature. Further research could be carried to find more precise relationship between them and the system could make the kurtosis and skewness high enough to improve the performance of system.

\bibliographystyle{IEEEtran}
\bibliography{refs}

%
%
%
%
%
%
%
%
%

\end{sloppy}
\end{document}